\documentclass[12pt]{article}

\usepackage{epsfig}

\usepackage{amsmath}
\usepackage{amssymb}
\usepackage{euscript}

\def\st{\hbox{}} 

\def\tbeta{\tilde{\beta}}

\newcommand{\Bmf}{\mathfrak{B}}

\begin{document}

\begin{titlepage}

\begin{flushright}
\bf IFJPAN-V-05-02
\end{flushright}

\vspace{5mm}
\begin{center}
  {\Large\bf%
    Studies of $\mu$-pair and $\pi$-pair production\\
    at the electron-positron low energy colliders$^{\star}$
}
\end{center}
\vspace{3mm}

\begin{center}
{\bf S. Jadach}

\vspace{1mm}
{\em Institute of Nuclear Physics, Academy of Sciences,\\
  ul. Radzikowskiego 152, 31-342 Cracow, Poland,}\\
\end{center}

\vspace{5mm}
\begin{abstract}
Predictions  for the radiative return with the muon pair and pion pair
final state from KKMC and PHOKHARA Monte Carlo programs
are compared and discussed.
The case of muon pairs is well understood,
especially of the initial state radiation (ISR), where three different
second order calculations agree very well.
The case of the final state radiation (FSR) requires more tests.
Matrix element in KKMC of the EEX type with the incomplete second order 
NLL corrections is not good enough for the radiative return at $Q^2<1$GeV 
with the precision requirement better than 1\%.
A method of extending the superior CEEX-type matrix element in KKMC to
the pion pair final state is described.
\end{abstract}

\vspace{4mm}
\begin{center}
\em Presented at the Epiphany Conference, Cracow, January 2005
\end{center}

\vspace{15mm}
\begin{flushleft}
{\bf IFJPAN-V-05-02
}
\end{flushleft}

\vspace{5mm}
\footnoterule
\noindent
{\footnotesize
$^{\star}$Supported in part by the EU grant MTKD-CT-2004-510126,
  in partnership with the CERN Physics Department
}

\end{titlepage}

\section{Introduction}
The aim of this contribution is to compare whatever the best
we have at hand for evaluation of the initial state radiation (ISR)
effect in the process $e^-e^+\to \mu^-\mu^+\gamma$
using KKMC \cite{Jadach:1999vf,Jadach:2000ir}
and PHOKHARA \cite{Rodrigo:2001kf,Czyz:2002np,Kuhn:2002xg}
Monte Carlo programs.
The above investigation will be partly extended
to the process $e^-e^+\to \pi^-\pi^+\gamma$.

\begin{figure}[!ht]
  \centering
  {\epsfig{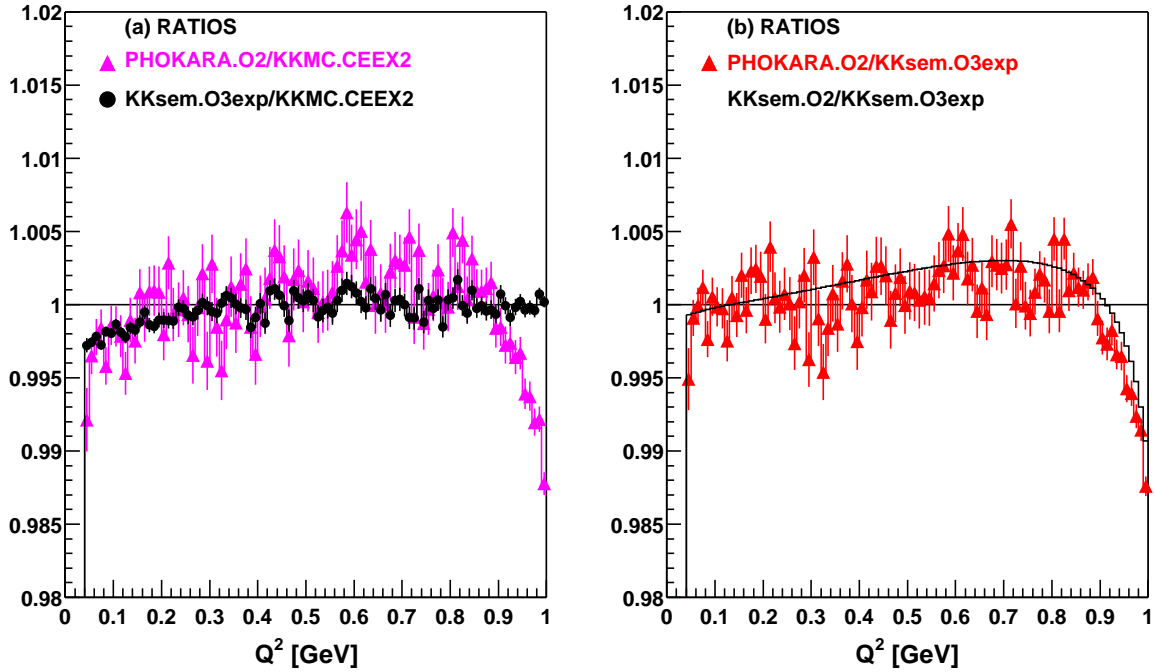}}
  \caption{\sf
    Mu-pair mass (square) spectrum in case of ISR only.
    $\sqrt{s}=1.01942$GeV.
    }
  \label{fig:1}
\end{figure}
\section{ISR in muon pair production}

Both KKMC and PHOKHARA programs are 
full scale MC event generators, which can provide for any
experimentally observable distribution.
We concentrate however, on the distribution of the squared mass spectrum 
$Q^2=s'$ of the muon pair,
because this distribution is relevant for the radiative return
measurements of $R(s)$,  and also because this particular distribution we
may compare with the classical semi-analytical calculations.
Here we shall also exploit the analytical formulas of 
ref.~\cite{Jadach:1991vz} (see also \cite{Jadach:1992aa}),
which implement analytical second order ISR calculation of ref.~\cite{BBVN:1986}
and third order leading-logarithmic (LL)
ISR calculation of refs.~\cite{Skrzypek:1992vk,Skrzypek:1991qs}.
The ISR formula of ref.~\cite{Jadach:1991vz} 
is provided by the KKsem facility of KKMC.
In the actual KKsem implementation we use version of the formula where
numerically negligible (at least at LEP energies, see ref.~\cite{Jadach:1991vz})
second order NNLL terms are neglected.

It is important to stress from the very beginning that authors of PHOKHARA
and KKMC use different terminology to describe Born level matrix element
and higher order matrix element.
I shall not try to unify terminology
or fully explain the differences, referring the reader
to original works, like refs.~\cite{Jadach:1999vf,Jadach:2000ir}
and \cite{Rodrigo:2001kf,Czyz:2002np,Kuhn:2002xg}.
Let me explain only very briefly the main differences.
The KKMC authors define Born as $e^+e^-\to f\bar{f}$ without any photon emission
and the radiative return is necessarily the first order process
with respect to such a Born level.
The leading-logarithmic (LL) corrections are of order $\alpha^n L^n$, 
where $n=1,2,3...\infty$ is the standard perturbative order,
while mass logarithm $L=\ln(s/m_e^2)$ 
is coming either from the virtual photon correction or the
phase space integration over the real
photon angle down to zero value.
The NLL and NNLL corrections are of order
$\alpha^n L^{n-1}$ and $\alpha^n L^{n-2}$ correspondingly.
Concerning mass terms, they are routinely neglected in KKMC for the electron
(except those which integrate up to a finite correction)
while an effort is made to keep all of them for the final state fermions, 
at least at the Born and the first order level.
KKMC implements several variants of the QED matrix elements, which feature
different level of higher order and mass term truncation.
PHOKHARA authors employ the leading-order (LO) as a name for the process
in which one (and only one) photon is emitted in the final state. 
They name as the next-to-leading-order (NLO)
their matrix element with the
one-loop corrections and the second real photon. 
This terminology may seem more adequate to discuss radiative return.
However, when trying to match the two terminologies
one has to pay attentions to the available phase
space of the first real photon.
Depending on whether the minimum emission angle
is imposed or not, one gets full factor $L$ or not, even at the LO.
This affects strongly the relative magnitude of higher order corrections 
with respect to LO or LL.
In this study we generally exclude from the considerations ``non-photonic''
corrections due to emission of additional lepton pair and vacuum polarization.

In Fig.\ref{fig:1} we compare results of KKMC and of PHOKHARA using the best
available ISR matrix element in both programs at $\sqrt{s}=1.01942$GeV.
In KKMC we use second order
matrix element with coherent exclusive exponentiation (CEEX)
described in refs.~\cite{Jadach:2000ir,Jadach:1998jb}.
The second order CEEX matrix element has complete
next-to-leading-logarithmic (NLL) contributions%
\footnote{For unpolarized beams, see discussion in ref.~\cite{Jadach:2000ir}
  related to eq.~(128) therein.}
and complete next-next-to-leading-logarithmic (NNLL) contributions.
The magnitude of NLL and NNLL corrections was also examined 
in a separate studies, see contribution of S.~Yost in these proceedings.
KKMC includes most of the third order LL contributions 
by the virtue of exponentiation%
\footnote{
  It is also known that exponentiation of 
  the YFS type sums up quite substantial
  part of third order LL, see refs.~\cite{Skrzypek:1992vk,Skrzypek:1991qs}}.
On the other hand, PHOKHARA implements complete second order ISR,
including complete NLL and NNLL corrections 
(i.e. singular corrections proportional to $\frac{\alpha}{\pi} m_e^2$
and $\frac{\alpha}{\pi} m_e^4$, which integrate to 
finite corrections of order $\frac{\alpha}{\pi}$, in the limit $m_e\to 0$). 
PHOKHARA does not resum (exponentiate) soft photon contributions to infinite order.
It is worth to stress that the two MC calculation, KKMC and PHOKHARA,
and semianalytical formula of KKsem of refs.~\cite{Jadach:1991vz,BBVN:1986}
represent set of {\em three completely independent} second order 
(using terminology of KKMC)
calculation of the ISR in every aspect of calculating QED matrix element
and integrating the phase space.

The main comparison of the ISR calculations is shown in Fig.\ref{fig:1}a,
where the distribution $d\sigma/dQ^2$ from
KKMC and KKsem agree very well, within 0.2\%, except very low $Q^2$
where they diverge by about 0.3\%
\footnote{ Note that similar comparison 
of KKMC and KKsem was done in ref.~\cite{Jadach:2000ir} 
for LEP energies. At the present lower energy $\sqrt{s}=1.01942$GeV 
subleading terms are, however, more important.},
while Fig.\ref{fig:1}a shows certain addition cross-check.
The reason for this discrepancy is not clear.
Neglected NNLL in KKsem are a viable candidate,
but to confirm this hypothesis one would need more tests.
In the same plot we see that PHOKHARA agrees well with KKsem at low $Q^2$
(aligning with KKsem) and
differs by about 0.25\% in the central region (we need higher statistics from
PHOKHARA to confirm this number) from both KKMC and KKSEM
and drops sharply at soft limit, high $Q^2$, because of lack
of soft photon resummation.
In order to understand quantitatively the effect of lack of exponentiation 
in PHOKHARA
we compare its result in Fig.\ref{fig:1}b with a variant of KKsem
in which we switch off exponentiation,
i.e. all terms beyond second order are truncated.
The smooth curve in Fig.\ref{fig:1}b representing result of this truncation
fits very well PHOKHARA result.
In particular, looking into this result, one may think that the deviation 
of PHOKHARA by 0.25\% in the central region is related 
to its neglect of the third order LL.
This conjecture needs more test to be confirmed.
\begin{figure}[!ht]
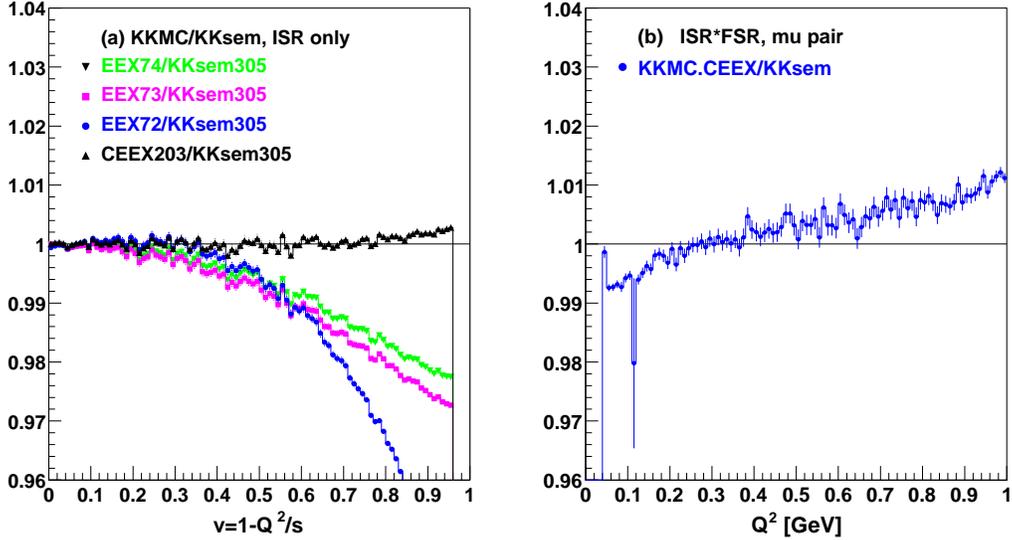

  \centering
  {\epsfig{file=./cFig3m2.eps.701M,width=70mm}}
  {\epsfig{file=./cFig3f1.eps.157M,width=70mm}}
  \caption{\sf
    Muon pair mass spectrum from KKMC and KKsem.
    }
  \label{fig:2}
\end{figure}

We summarize on the results of Fig.\ref{fig:1}
that KKMC with the second order CEEX matrix element,
PHOKHARA with its second order matrix element and KKsem implementing
second order analytical calculation agree very well,
within the expected range and the pattern of the discrepancies
seems to be understood.

In KKMC there is another more primitive QED matrix element denoted as EEX,
see ref.~\cite{Jadach:2000ir} for its full description, 
which follows closely the classical Yennie-Frautschi-Suura (YFS) exponentiation
scheme and its implementation is limited to first order plus second order LL.
In the second order EEX matrix element (contrary to CEEX)
the NLL corrections are incomplete.
(On the other hand EEX third order LL is complete, while in CEEX it is incomplete.).
For technical and historical reasons,
see discussion below, EEX type matrix element is used
for the production of low energy hadronic final states, for example for pion pair.
It is, therefore, important to check how good it is compared to 
KKMC with more complete coherent exclusive  exponentiation (CEEX) matrix element.
This is done for the muon final state in  Figs.~\ref{fig:2}.
(CEEX is not yet available for $\pi$-pairs).
In Fig.~\ref{fig:2}a we see results from KKMC CEEX and several variants of EEX.
We are actually plotting $d\sigma/dQ^2$,
dividing all results by KKsem of ref.~\cite{Jadach:1991vz}, 
the same as in previous Fig.\ref{fig:1}.
The curves marked EEX72 represent exponentiated EEX matrix element
based on complete first order,
while EEX73 and EEX74 include also complete
second and third LL, while CEEX203%
\footnote{Indices 203, 74 etc. follow numbering of MC weights 
in ref.~\cite{Jadach:1999vf}.}
is the same as in Fig.~\ref{fig:1}.
As we see, at low $Q^2$, that is for the hard photon emission,
results of EEX matrix elements
depart from other more complete results by up to 3\%!
In the $\rho$ region it is different from the KKsem, CEEX KKMC
by about 1\%.
The above result is also consistent with what we have seen in Fig.~\ref{fig:1}.
{\em EEX is therefore not well suited for the use in the high precision
measurements of $R(s)$ using radiative return below $Q^2$=1GeV.}
This result is not very much surprising,
as EEX of KKMC has incomplete second order NLL.
The observed effect at the low $Q^2$ is a little bit bigger then what we expected.
We have therefore done certain additional tests.
We have split EEX results into three components, $\tbeta_i,\; i=0,1,2$,
compared each of them with analytical result of table~I of
ref.~\cite{Jadach:2000ir}, also at $\sqrt{s}=10$GeV.
We do not show results of these tests here, but the overall pattern of discrepancies
seems to be consistent with NLL class of corrections.
This additional test indicates also that
the main source of the problem is an approximate double real
emission matrix element in EEX and not the incomplete virtual corrections.
In particular we have included in these tests the complete 
NLL contribution in $\tbeta_0$ and $\tbeta_0$. This did not help!
The whole discrepancy seems to result from the use of the LL-approximate
matrix element for the double real photon emission in EEX.
The above observation is consistent with the older
tests in ref.~\cite{Jadach:2000ir} at LEP energies.

\begin{figure}[!ht]
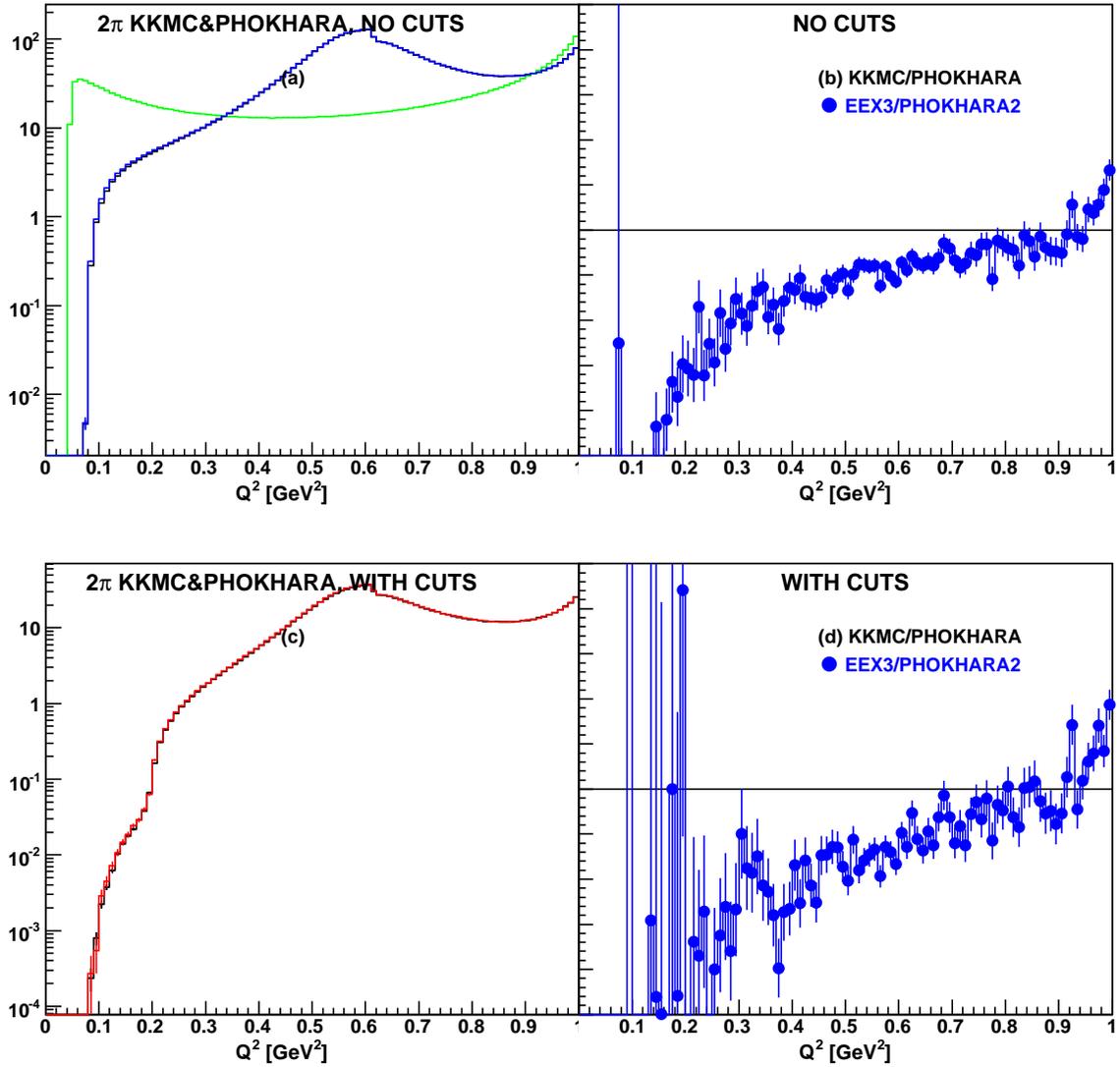

  \centering
  {\epsfig{file=./cFig3a1.eps.960M,width=160mm}}
  \\
  {\epsfig{file=./cFig3a2.eps.960M,width=160mm}}
  \caption{\sf
    Comparison of PHOKHARA NLO and KKMC with EEX matrix element. ISR only.
    }
  \label{fig:3}
\end{figure}
\subsection{Muon pair, ISR+FSR}

Let us not include FSR in the game, again for the muon pair final state.
In fact, at low $Q^2$ the rate of muon pair
in radiative return is higher than of $\pi$ pairs,
hence $d\sigma/dQ^2$ of muons can be used as a reference distribution
for measuring $R(s)$.
It is therefore worth to test FSR in KKMC
and to check result for $d\sigma/dQ^2$ once again.
In Fig.~\ref{fig:2}b we show result from KKMC for 
second order CEEX matrix element
in which we include ISR, FSR and its interference.
We compare MC result with the semianalytical result of KKsem in which
with the same ISR radiative function of ref.~\cite{Jadach:1991vz}.
The FSR distribution of ref.~\cite{Jadach:2000ir}
features incomplete NLL in KKsem, so it is definitely inferior with respect
to ISR counterpart -- the complete list of the FSR radiative corrections
in KKsem can be found in Table~II in ref.~\cite{Jadach:2000ir}.
This above semianalytical formula also misses the interference of ISR and FSR,
which in first order is zero in the inclusive $d\sigma/dQ^2$
so this omission does not harm.
In Fig.~\ref{fig:2}a we see the ratio of the corresponding results 
from KKMC and KKsem.
(NB. PHOKHARA is able to provide result with FSR for muon pairs in the LO,
and it would be interesting to include it in the comparison.)
This result is rather preliminary and has to be checked.
In any case, the agreement better than 1\% found all over the $dQ^2$ range is
quite satisfactory as a starting point for further investigation%
\footnote{In one bin we see trace of large weight fluctuation which is probably
  due to rounding errors. This result was obtained using weighted events.
  For the MC run with weight-one events this effect would disappear. 
  Such numerical instabilities need further investigation.}

\section{ISR for $\pi^+\pi^-$ pair production}

Let us now switch to low $Q^2$ $\pi$-pair state produced at the
radiative return process.
In Fig.~\ref{fig:3} we compare KKMC with the EEX style matrix element on one hand
with PHOKHARA second order (marked as PHOKHARA2) on the other hand.
The EEX matrix element is the default one in KKMC, with first order exponentiation,
the completed second and third order LL (EEX74).
We limit ourselves to ISR only.
The results of Figs.~\ref{fig:3}a-b are obtained without any cutoffs.
In Fig.~\ref{fig:3}a we show the actual distributions,
including also the distribution for the muon pairs.
We see that for $Q^2<0.33$ the muon-pair cross section 
is bigger than that of $\pi$-pair.
The ratio PHOKHARA/KKMC is not so well understood as the the analogous results
for the muon pair shown in the previous section.
The discrepancy at high $Q^2$ we attribute to lack of exponentiation
in PHOKHARA%
\footnote{This does not hinder practical applications of PHOKHARA
for radiative return measurements, which concentrate at lower $Q^2$.}
while another larger discrepancy at low $Q^2$ is most likely due 
to incompleteness of second order NLL in EEX
matrix element on KKMC,
and it corresponds to deviation which was already seen in Fig.~\ref{fig:2}a.
In Fig.~\ref{fig:3}c-d we show the analogous results for relatively mild cut
on photon momentum, where photon is defined 
as a ``missing four momentum''
calculated knowing pion momenta and beam momenta.
We ask for the momentum of such a ``collective unseen photon''
to be directed below $15^\circ$ from the beam and to have at least $10MeV$
of energy.
For each $\pi$ we require that it is situated in wide angles,
eg. separated by more than $40^\circ$ from each beam.
Results in Fig.~\ref{fig:3}c-d look quite similar,
except that the discrepancy between
PHOKHARA and EEX KKMC is bigger 
(we need better statistics from PHOKHARA to see it more clearly).
This can be attributed to the fact that the leading logarithm $L$
due to real emission
is diminished by the cut on the photon angle with respect to beams.

\subsection{How to extend CEEX ISR to hadronic final states?}
In the following we show that the superior CEEX ISR matrix element can be extended
to hadronic final states at low $Q^2$, like pion pair.
This can be done provided we have some decent modeling of the hadronic
final state in terms of the corresponding formfactor.
In view of its practical importance, let us elaborate on this point.

In CEEX Born amplitude for $ee\to\mu\mu$ is defined as a 
four-spinor tensor
\def\st{\hbox{}} 
\begin{equation}
  \label{eq:born-def}
  \begin{split}
     &\Bmf\left(\st^{p}_{\lambda}; X \right)=
      \Bmf\left(  \st^{p_a}_{\lambda_a} \st^{p_b}_{\lambda_b}
                  \st^{p_c}_{\lambda_c} \st^{p_d}_{\lambda_d}; X \right)=
      \Bmf\left[  \st^{p_b}_{\lambda_b} \st^{p_a}_{\lambda_a}\right]
          \left[  \st^{p_c}_{\lambda_c} \st^{p_d}_{\lambda_d}\right]\!(X)=
      \Bmf_{[ba][cd]}(X)=
\\  &\qquad\qquad
      =ie^2 \sum_{B=\gamma,Z} \Pi^{\mu\nu}_B(X)\; (G^{B}_{e,\mu})_{[ba]}\; (G^{B}_{f,\nu})_{[cd]}\; H_B
      =\sum_{B=\gamma,Z} \Bmf^B_{[bc][cd]}(X),
\\
     &(G^{B}_{e,\mu})_{[ba]} \equiv \bar{v}(p_b,\lambda_b) G^{B}_{e,\mu} u(p_a,\lambda_a),\;\;
      (G^{B}_{f,\mu})_{[cd]} \equiv \bar{u}(p_c,\lambda_c) G^{B}_{f,\mu} v(p_d,\lambda_d),
\\
     &G^{B}_{e,\mu} = \gamma_\mu \sum_{\lambda=\pm} \omega_\lambda g^{B,e}_\lambda,\quad
      G^{B}_{f,\mu} = \gamma_\mu \sum_{\lambda=\pm} \omega_\lambda g^{B,f}_\lambda,\quad
      \omega_\lambda = {1\over 2}(1+\lambda\gamma_5),
\\
     &\Pi^{\mu\nu}_B(X) = { g^{\mu\nu} \over X^2 - {M_{B}}^2 +i\Gamma_{B} {X^2 / M_{B}} },
  \end{split}
\end{equation}
and it enters as a basic building block in every spin amplitude in the CEEX scheme,
with arbitrary number of photons.
See eq.~(43) in ref.~\cite{Jadach:2000ir} for notation.
The above Born is calculated using Chisholm identity and replaced with the
bi-spinor objects of the Kleiss-Stirling method.

In case of hadronic final state the structure of the Born amplitude is
\begin{equation}
 \Bmf^\mu_{[ba]}(X) J_\mu(X, q_i),\quad J_\mu(X, q_i) X^\mu=0,
\end{equation}
where $q_i$ are momenta of the final state hadrons, $X=\sum q_i$, and
\begin{equation}
  \begin{split}
    &\Bmf^\nu_{[ba]}(X)
      =ie^2 \sum_{B=\gamma,Z} \; H_B\; (G^{B}_{e,\mu})_{[ba]}\; 
                    \Pi^{\mu\nu}_B(X)
      =\sum_{B=\gamma,Z} \Bmf^{B\mu}_{[bc]}(X).
  \end{split}
\end{equation}
In the rest frame of $X$ one has $J^\mu=(0,\vec J )$
and we may split $J$ into difference of the two massless four-vectors
$J^\mu = J^\mu_+ - J^\mu_-$.
In the arbitrary reference frame the above prescription extends as follows%
\footnote{The author would like to thank J. Kuehn for suggesting him 
this solution.}
\begin{equation}
  J^\mu_\pm = \frac{1}{2\sqrt{X^2}} 
  ( \sqrt{-J^2}\; X^\mu \pm \sqrt{X^2}\; J^\mu ).
\end{equation}
Each of the two corresponding components in $\Bmf$ can be expressed
in terms of the  of the Kleiss-Stirling bi-spinors
$s_\pm(p_1,p_2)$, see eqs.~(A4-A6) in ref.~\cite{Jadach:2000ir}.
This can be done using completeness relation for
$\bar{v}(p_b,\lambda_b)\; \gamma_\nu J^\nu_\pm\;  u(p_a,\lambda_a)  $
taking advantage of $J^2_\pm=0$.

Note that the above CEEX implementation  requires that we parametrize 
the production amplitude of the each final hadronic state one by one 
in terms of the formfactors in a completely exclusive manner.
However, this is necessary anyway for good phenomenological
description of these often resonant low energy hadronic states.
We conclude that CEEX can be used for ISR for low energy hadron production.
The question is only how much programming it will be and who will do it.

\subsection{Conclusions}
The case of radiative return with the muon pair final state
is well understood, especially for the ISR where three different
second order calculations agree very well.
The case with FSR requires more tests.
Matrix element EEX of KKMC with the incomplete second order 
NLL is not good in the radiative return at $Q^2<1GeV$ 
for precision requirement better than 1\%.
Method of porting CEEX matrix element of KKMC to pion pair final state
is outlined.

\vspace{8mm}
{\bf\large Acknowledgments}\\
This work is partly supported by TARI Contract No.~RII3-CT-2004-506078.
The author would like to thank J. Kuehn and A. Denig for useful discussion.
Warm hospitality at TTP Karlsruhe University and INFN Frascati, where part of
this work was done is also acknowledged.


\begin{thebibliography}{10}

\bibitem{Jadach:1999vf}
S.~Jadach, B.~F.~L. Ward, and Z.~Was, {\em Comput. Phys. Commun.} {\bf 130}
  (2000) 260--325,
\href{http://arXiv.org/abs/hep-ph/9912214}{{\tt hep-ph/9912214}}.

\bibitem{Jadach:2000ir}
S.~Jadach, B.~F.~L. Ward, and Z.~Was, {\em Phys. Rev.} {\bf D63} (2001) 113009,
\href{http://arXiv.org/abs/hep-ph/0006359}{{\tt hep-ph/0006359}}.

\bibitem{Rodrigo:2001kf}
G.~Rodrigo, H.~Czyz, J.~H. Kuhn, and M.~Szopa, {\em Eur. Phys. J.} {\bf C24}
  (2002) 71--82,
\href{http://arXiv.org/abs/hep-ph/0112184}{{\tt hep-ph/0112184}}.

\bibitem{Czyz:2002np}
H.~Czyz, A.~Grzelinska, J.~H. Kuhn, and G.~Rodrigo, {\em Eur. Phys. J.} {\bf
  C27} (2003) 563--575,
\href{http://www.arXiv.org/abs/hep-ph/0212225}{{\tt hep-ph/0212225}}.

\bibitem{Kuhn:2002xg}
J.~H. Kuhn and G.~Rodrigo, {\em Eur. Phys. J.} {\bf C25} (2002) 215--222,
\href{http://www.arXiv.org/abs/hep-ph/0204283}{{\tt hep-ph/0204283}}.

\bibitem{Jadach:1991vz}
S.~Jadach, M.~Skrzypek, and B.~F.~L. Ward, {\em Phys. Lett.} {\bf B257} (1991)
173--178.

\bibitem{Jadach:1992aa}
S.~Jadach, M.~Skrzypek, and M.~Martinez, {\em Phys. Lett.} {\bf B280} (1992)
129--136.

\bibitem{BBVN:1986}
F.~Berends, G.~Burgers, and W.~V. Neerven, {\em Phys. Lett.} {\bf 177} (1986)
  1191.

\bibitem{Skrzypek:1992vk}
M.~Skrzypek, {\em Acta Phys. Polon.} {\bf B23} (1992)
135--172.

\bibitem{Skrzypek:1991qs}
M.~Skrzypek and S.~Jadach, {\em Z. Phys.} {\bf C49} (1991)
577--584.

\bibitem{Jadach:1998jb}
S.~Jadach, B.~F.~L. Ward, and Z.~Was, {\em Phys. Lett.} {\bf B449} (1999)
  97--108,
\href{http://www.arXiv.org/abs/hep-ph/9905453}{{\tt hep-ph/9905453}}.

\end{thebibliography}

\providecommand{\href}[2]{#2}

\end{document}